\documentclass[aps,pra,twocolumn,superscriptaddress,floatfix]{revtex4-2}
\usepackage{amssymb}
\usepackage{graphicx}
\usepackage{dcolumn}
\usepackage{bm}
\usepackage{amsmath}
\usepackage{subfigure}
\usepackage{float}
\usepackage{color}
\usepackage[colorlinks=true,citecolor=blue]{hyperref}
\usepackage{multirow}

\begin{document}

%\preprint{APS/123-QED}

\title{Feedback Intensity Equalization Algorithm for Multi-Spots Holographic Tweezer}

\author{Shaoxiong Wang}
\thanks{These authors contributed equally to this work.}
\affiliation{State Key Laboratory of Quantum Optics Technologies and Devices, Institute of Opto-Electronics, Shanxi University, Taiyuan 030006, China}

\author{Yifei Hu}
\thanks{These authors contributed equally to this work.}
\affiliation{State Key Laboratory of Quantum Optics Technologies and Devices, Institute of Opto-Electronics, Shanxi University, Taiyuan 030006, China}

\author{Yaoting Zhou}
\affiliation{State Key Laboratory of Quantum Optics Technologies and Devices, Institute of Opto-Electronics, Shanxi University, Taiyuan 030006, China}

\author{Peng Lan}
\affiliation{School of Physics and Electronics Engineering, Shanxi University, Taiyuan, Shanxi 030006, China}

\author{Zhongxiao Xu}%
\email{xuzhongxiao@sxu.edu.cn}
\affiliation{State Key Laboratory of Quantum Optics Technologies and Devices, Institute of Opto-Electronics, Shanxi University, Taiyuan 030006, China}
\affiliation{Collaborative Innovation Center of Extreme Optics, Shanxi University, Taiyuan 030006, China}

\date{\today}% It is always \today, today,

\begin{abstract}
High degree of adjustability enables the holographic tweezer array a versatile platform for creating an arbitrary geometrical atomic array. In holographic tweezer array experiments, an optical tweezer generated by a spatial light modulator (SLM) usually is used as a static tweezer array. However, the alternating current (AC) stark effect generally induces the intensity difference of traps in terms of different light shifts. So, intensity equalization is an essential prerequisite for preparing a many-body system with individually controlled atoms. Here, we report an intensity equalization algorithm. In particular, we observe the non-uniformity of the tweezer  array is below 1.1$\%$ when the array size is larger than 1000. Our analysis shows that by optimizing the hardware performance of the optical system, this uniformity could be further improved. Our work offers the opportunities for large-scale quantum computation and simulation with reconfigurable atom arrays.
\end{abstract}

%\keywords{}

\maketitle

\section{Introduction}
The holographic tweezer array formed by SLM is a technique that uses interference and diffraction principles to reconstruct the desired optical arrays \cite {Naoya,Wang}. There are several protocols to produce the tweezer array, such as, by digital micro-mirror devices (DMD) \cite {Zhang}, by Talbot effect \cite {Zhang:22}, by micro-lens\cite{Xixi2020SubwavelengthIA}, by metasurface \cite{article}. Compared to the SLM, the diffraction efficiencies of DMD and meta-surface are limited, while the arbitrary editing capability is impossible for Talbot and micro-lens array. In contrast, the phase pattern of SLM can be set on demand \cite{Dholakia2011ShapingTF,Alexander L}, which makes it flexible for producing arbitrary arrays and correcting the aberration 
 \cite{Bruce2014FeedbackenhancedAF,Dixon2017UsingBF,David2013Calibration,Zhao}. Therefore, the SLM has become a high-efficiency equipment for generating optical tweezer arrays \cite{2014Single,2014Stable}. 

Recent decade has witnessed great progresses in holographic tweezer arrays of neutral atoms. A key idea is outlined as follows: When the tweezer waist is small enough, the collisional-blocked effect leads to the fact that one atom at most can be trapped in the tweezer. Combined with the rearrangement technique, two- and three-dimensional defect-free atomic arrays have been realized \cite{2022In,Joannopoulos}. This reconfigurable platform opens a new avenue to large-scale quantum computation \cite{Saffman} and simulation \cite{Andrew,Bernien}, and quantum metrology \cite{0Quantum}. 

Concretely, the tweezer arrays made by SLM are used for static tweezer that plays the role of trapping the initial atomic array for sorting and holding the sorted atomic array \cite{Kim}. In fact, the trap depth is proportional to the trap intensity,  and an intensity equalized trap array is necessary \cite{Kim} to prepare an ideal atomic array. Otherwise, an inhomogeneous trap depth in the tweezer array will result in the difference of the loss probability of the trapped atom as well as the different light shifts. Here we report a new scheme of the intensity equalization enabled by feedback weighted-Gerchberg-Saxton (GSW) algorithm \cite{GanLight,Honghao}. Through the imaging of the generated array, we create the corrected target and feed it back to the GSW algorithm. By leveraging the iterative feedback process, the non-uniformity of the tweezer array can be reduced significantly. The results show that the non-uniformity can be below 1.1\% when the array size is over 1000.  

\maketitle
\section{Principle of Feedback GSW}

\begin{figure*}
\centering
\includegraphics[width=0.9\textwidth]{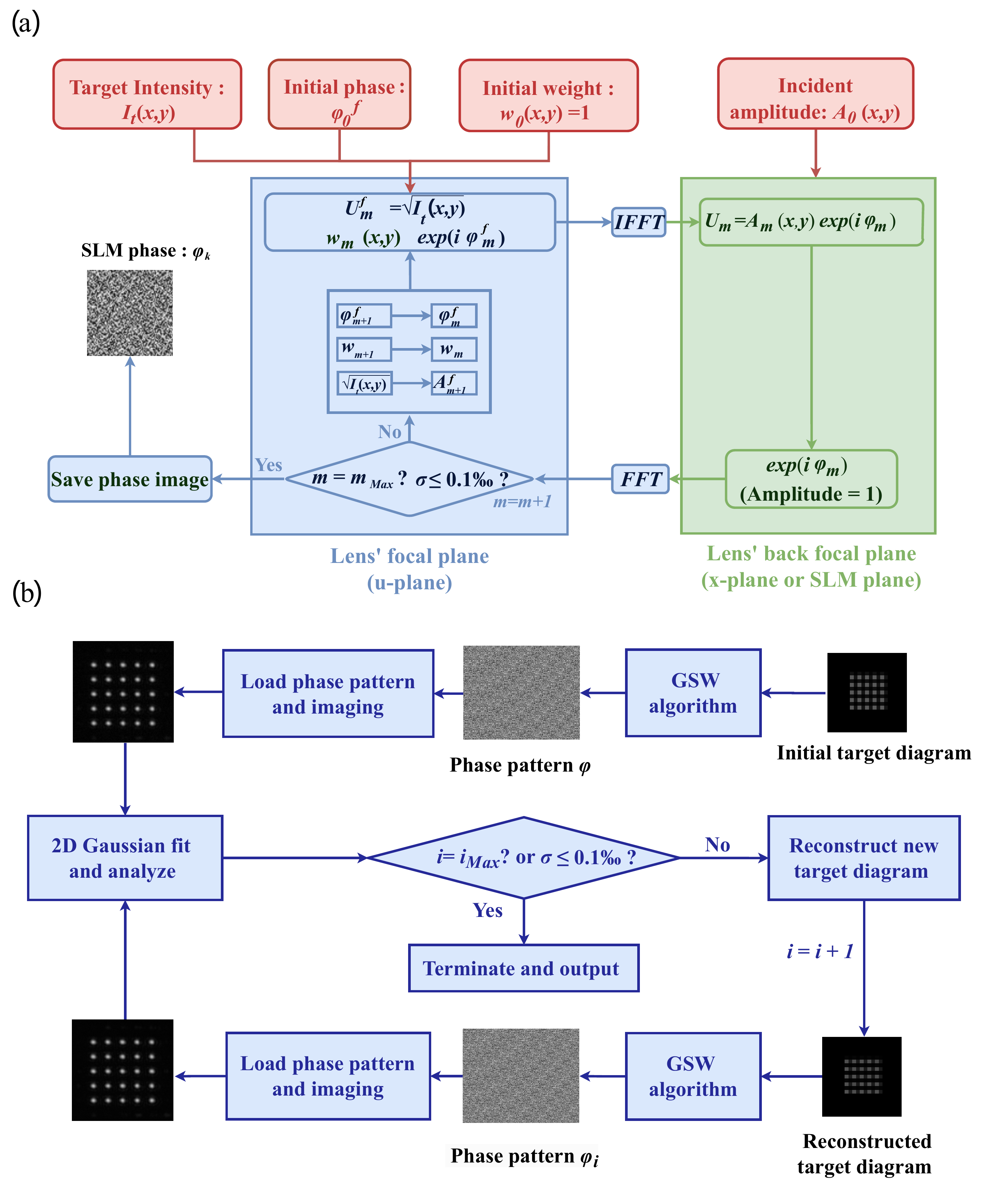}
\caption{\label{Fig1} (a) The weighted-Gerchberg–Saxton (GSW) algorithm. (b) Feedback algorithm used for reducing the non-uniformity of holographic optical tweezer array. The input target diagram is an 8-bit deep map, and the initial brightness of spots is set at 128. By using the GSW algorithm, a corresponding phase pattern $\varphi$ will be produced. After loading the phase pattern to the SLM, an actual tweezer array will be imaged by a CCD. The intensity information of the array will be achieved by 2D Gaussian fit and analysis. Then the uniformity and cycle number will be judged. If the uniformity or cycle number reaches the set value, the cycle ends. Otherwise, a corrected target will be reconstructed and induced into the feedback loop}
\end{figure*}

The GSW algorithm \cite{GanLight} is the core of our approach. The specific process of the GSW algorithm and feedback is shown in Fig.\ref{Fig1}.The algorithm is based on Fourier optics theory and iteratively adjusts the phase distribution on a SLM to achieve the desired intensity distribution on the focal plane. The GSW algorithm optimizes the uniformity and quality of the holographic optical tweezer array by introducing weighted factors. We first generate a desired target diagram. The SLM is placed at the back focal plane (x-plane) of the lens. The reflected wavefront is shaped by a computer-generated hologram (CGH) \cite{2017Vortex} displayed on the SLM, and then forms the tweezer array at the focal plane (u-plane) of the lens. 

The electric component intensity distribution in u-plane can be expressed as $U_{0}^{f} =\sqrt{I_{t}(x,y) } \times  e^{i\times {\varphi _{0}^{f} } } $, where $I_{t} (x,y) $ is the desired intensity distribution, and $\varphi _{0}^{f} $ is the initial random phase we give for iteration. 

According to the imaging principle of Fourier optics, the pattern in the x-plane can be calculated to be $A_{m}(x,y)  \times  e^{i\times\varphi _{m}  } $. Due to the phase-only spatial light modulator, here we assign the amplitude $A_{m}(x,y)$ to 1 and keep the phase component $  e^{i\times\varphi _{m}  } $ for fast Fourier-transform(FFT). A new target graph $U_{m}^{f} (x,y)$ can be obtained by this FFT. So we could get the intensity distribution $I_{m}^{f} (x,y)=\mid FFT(e^{i\times\varphi  _{m} })\mid ^{2} $. To approach the desired target, a weighted factor is calculated by the expression below:

\begin{equation}\label{Eq1}
    w_m (x,y)= \frac{I_t (x,y)}{I_m^f (x,y)}\times w_{m-1} (x,y)
\end{equation}

And the modified target will be $\sqrt{I_{t}(x,y) } \times w_{m}(x,y) \times e^{i\times \varphi _{m+1} } $.

After running many iterations inside the algorithm, the calculated target will be close to the desired target we input. 

In practice, a grating phase needs to be added to shift the target away from the zero-th order. And due to the imperfect optical system, the achieved target is not uniform as the calculated result. To correct this non-uniformity caused by the optical system, we apply another feedback scheme. The basic principle is modifying the input target of GSW according to the non-uniformity target achieved by the GSW algorithm.
This so-called feedback GSW algorithm begins with imaging the i-th GSW algorithm and analyzing the non-uniformity by 2D Gaussian fit of each spot.
The non-uniformity (standard deviation) is calculated respectively by
\begin{equation}\label{Eq2}
    \sigma =\sqrt{\frac{1}{K} {\textstyle \sum_{k=1}^{K}}\left [ I_{k,i} -\bar{I}_i   \right  ]^{2}    }  
\end{equation}

where K represents the total number of optical holographic tweezers, $I_{k,i}$ is the i-th intensity of each spot, and $\bar{I_i} $ is the average intensity of 2D Gaussian fit intensities in the i-th iteration. The (i+1)-th intensity is calculated as

\begin{equation}\label{Eq3}
    I_{re,i+1} = \frac{I_{t,i}\times \bar{I_i}}{\bar{I_i}+(I_{k,i}-\bar{I_i})\times G  }   
\end{equation}

where $I_{re,i+1} $ is the reconstructed target intensity for (i+1)-th iteration, $I_{t,i}$ is the target intensity of last loop. G is an adjustable gain factor. Here we apply a constant number as the gain factor. It turns out that the best performance is achieved when $G=0.7$. Smaller value can slow down the convergence of the algorithm, while bigger value can cause problems with over-correction. By this method, the input target of the GSW algorithm is modified according to the achieved target. And the non-uniformity caused by the imperfect optical system can be optimized.

\maketitle
\section{Imaging and analyzing}

\begin{figure}
\centering
	\includegraphics[width=0.5\textwidth]{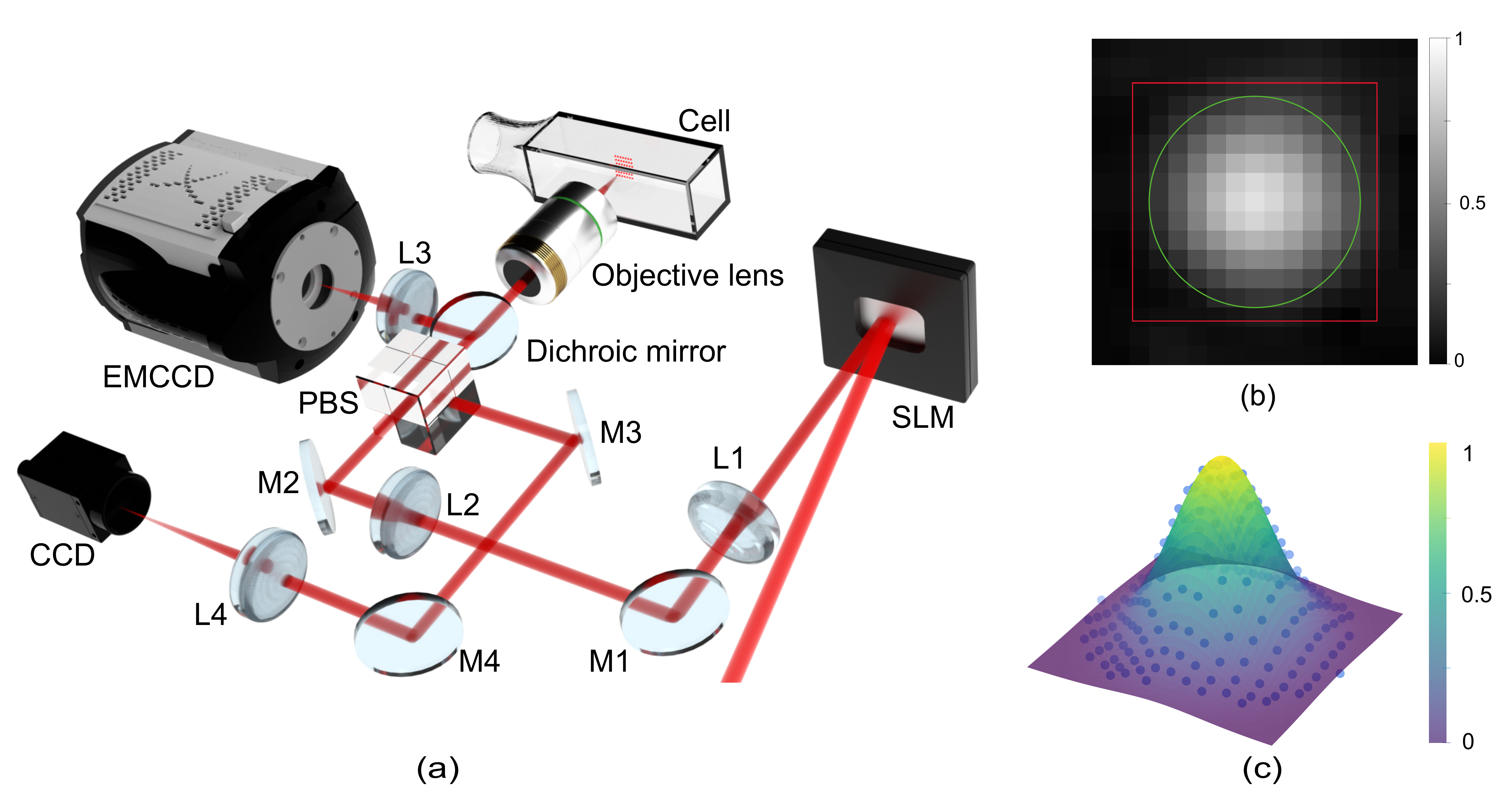}
\caption{(a) Experimental setup. L1: f=500 mm, L2: f=300 mm, L3: tube lens f=200 mm, L4: f=250 mm, M1-M4: High reflectivity (HR) mirrors (b) Tweezer spot details acquired by CCD. (c) 2D intensity distribution (Blue spots) and 2D Gaussian fit}
\label{fig2}
\end{figure}

The experimental setup is shown in Fig.\ref{fig2}(a). A collimated 852 nm laser beam illuminates the SLM (LCOS-SLM 15213-02, Hamamatsu Photonics) with a diameter of 10 mm. Then the wavefront modulated beam passes through a Relay lens system (L1 and L2) and the phase pattern is transferred to the back pupil plane of the objective lens (Mitutoyo G Plan APO 50X). We insert a polarizing beam splitter (PBS) after L2 to pick up a small amount of light for intensity monitoring by a charge coupled device (CCD) camera (JHEM506GM). The most of power goes through an objective lens and then generates the tweezer array in vacuum chamber for trapping the atomic array. The backward fluorescence is picked up by a dichroic mirror and collected by an electron-multiplying CCD (EMCCD). The spot diameter is $\sim$1 $\mu$m in the vacuum chamber and is $\sim$50 $\mu$m on CCD.

The expose duration of CCD is set as 10 $\mu$s. For the system preparation process, the image from CCD will be used for intensity equalization, while for the trapping process, the image from CCD will be used for pointing locking.

Since the CCD is an 8-bit deep camera, we can observe the saw-tooth shape from the 2D intensity distribution drawing in Fig.\ref{fig2} (c).  The pixel size of the CCD we are using is  3.45 $\mu$m $\times$3.45 $\mu$m. So, the focus will only take $\sim $ 15 pixels and it is impossible to neglect the saw-tooth shapes (detailed in Fig.\ref{fig2} (b)). To avoid the effect of saw-tooth shapes on the estimation of intensity equalization, we use the two-dimensional Gaussian fit method to fit the tweezer intensity.

The fit formula is:
\begin{equation}\label{Eq4}
    f(x,y)=Ae^{-[a(x-x_0)^2+2b(x-x_0)(y-y_0)+c(y-y_0)^2]}+D
\end{equation}
where
\begin{align}
 a&=\frac{\cos^2(\theta)}{2\sigma_x^2}+\frac{\sin^2(\theta)}{2\sigma_y^2},\\ \label{Eq5}
 b&=-\frac{\sin(2\theta)}{4\sigma_x^2}+\frac{\sin(2\theta)}{4\sigma_y^2},\\
 c&=\frac{\sin^2(\theta)}{2\sigma_x^2}+\frac{\cos^2(\theta)}{2\sigma_y^2}.
\end{align}

Here we use the general two-dimensional Gaussian fit to determine the intensity, beam waist, and ellipticity simutanously. In Eq.\ref{Eq4}, $A$ is the amplitude of the Gaussian fit function, representing the square root of intensity. And $x_0$, $y_0$ are the center coordinates of the Gaussian function. $D$ is the offset of the Gaussian function, which represents the intensity of the background. In Eq.5, $\sigma_x$ and $\sigma_y$ are the standard deviation of the Gaussian function in the x and y directions, which represents the width of the waist in the x and y directions respectively. $\theta$ is the rotation angle of the Gaussian fit function which can reflect the aberration problem. 

We evaluate the time cost for producing a 32$\times$32 intensity equalized array. From inducing a target input to producing the first phase pattern, the GSW algorithm runs 20 cycles and takes less than 1 ms. Then the feedback process runs 10 cycles. Each feedback cycle contains several steps including finding out the position of the spots, 2D Gaussian fit, intensity analysis, producing the new target, inducing the next GSW algorithm, and producing a new phase pattern. Each feedback process takes $ \sim $ 23 s. Hence, a total time of 4 min to 5 min is needed for producing an intensity equalized tweezer array with a size larger than 1000. 

\maketitle
\section{Results}

Several algorithms can realize the intensity equalization. We test the non-uniformity of a 32$\times$32 tweezer array produced by different algorithms. According to the Gerchberg-Saxton (GS) 
 \cite{2004Interactive}, Generalized Adaptive Additive (GAA) \cite{2002Dynamic} and GSW algorithms, the non-uniformity produced by the GSW algorithm is the least. Table \ref{tab1} is the test result of different strategies.

The performance of different strategies is quantified by non-uniformity (standard deviation,$\sigma$). We use Eq.\ref{Eq2}. to character the non-uniformity of the array, where K represents the total number of optical holographic tweezers,$I_{k,i}$ is the i-th intensity of each spot.

\begin{table}[htbp]
 \centering

  \caption{Intensity equalization result. After running 20 times, experimental performance of 32$\times$32 tweezers under different algorithms}\label{tab1}

  \renewcommand\arraystretch{1.5} 
\begin{tabular}{|c|cc|cc|}
\hline
\multirow{2}{*}{Algorithm} & \multicolumn{2}{c|}{Simulation} & \multicolumn{2}{c|}{Experiment} \\ \cline{2-5} 
                           & \multicolumn{2}{c|}{${\sigma}$(\%)}       & \multicolumn{2}{c|}{${\sigma}$(\%)}       \\ \hline
GS                         & \multicolumn{2}{c|}{18.18}      & \multicolumn{2}{c|}{24.84}      \\ \hline
GAA                        & \multicolumn{2}{c|}{14.81}       & \multicolumn{2}{c|}{24.69}       \\ \hline
GSW                        & \multicolumn{2}{c|}{0.44}       & \multicolumn{2}{c|}{11.007}       \\ \hline
F-GSW                      & \multicolumn{2}{c|}{-}          & \multicolumn{2}{c|}{1.006}       \\ \hline
\end{tabular}
\end{table}

The experimental results show that for a 32$\times$32 array, the non-uniformity of simulated GSW is 0.44$\%$ and it is 11.007$\%$ when directly loading the phase pattern on the SLM, which is due to the imperfect optical system. After our feedback optimization, the non-uniformity can be reduced to 1.006$\%$. Compared to the theoretical limit of 0.44$\%$ calculated by the GSW algorithm, it is only 0.566$\%$ difference after our feedback optimization, which means most of the non-uniformity caused by the optical system has been corrected.
\begin{figure}[ht!]
\centering\includegraphics[width=1\columnwidth]{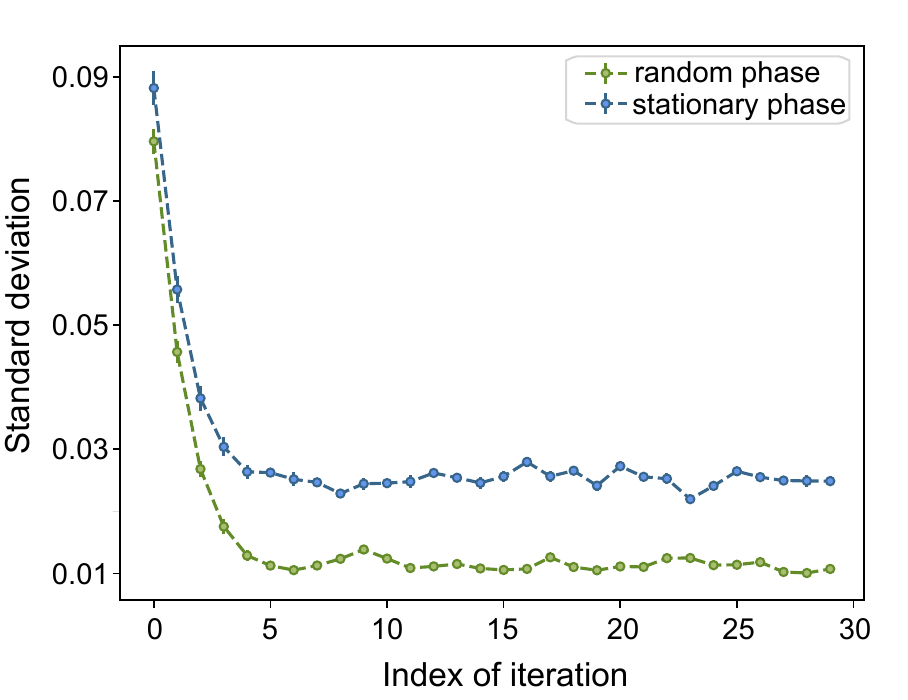}
\caption{Comparison results of stationary phase and random phase methods. The green dots are the results of experiments that fixed the initial phase. The blue dots are the results of experiments with random initial phase. The error bar is one standard variance and is calculated by 10 experimental runs}\label{fig3}
\end{figure}

For the GSW algorithm, a random initial phase is given for iteration. This leads to the instability of the intensity equalization results. In the GSW algorithm we using, if the initial phase is fixed each time, then the phase diagram produced by the iteration is the same. Therefore, in the calculation process, we first use random initial phases for several GSW iterations and take the initial phase corresponding to the best uniformity as the initial phase of the feedback iteration. We call this the stationary phase method. The purpose of this process is to select the most effective phase to improve the uniformity of the holographic tweezers array rather than relying on more iterations to optimize the experimental results.We compare the non-uniformity results of the random initial phase and fixed initial phase, and the experimental results are shown in Fig.\ref{fig3}. The non-uniformity of stationary initial phase method is about 1.5 percentage points lower than that of the random phase method. And the error bar is much smaller. The error bars are calculated by 10 experiment runs. 

\begin{figure}[ht!]
\centering\includegraphics[width=1\columnwidth]{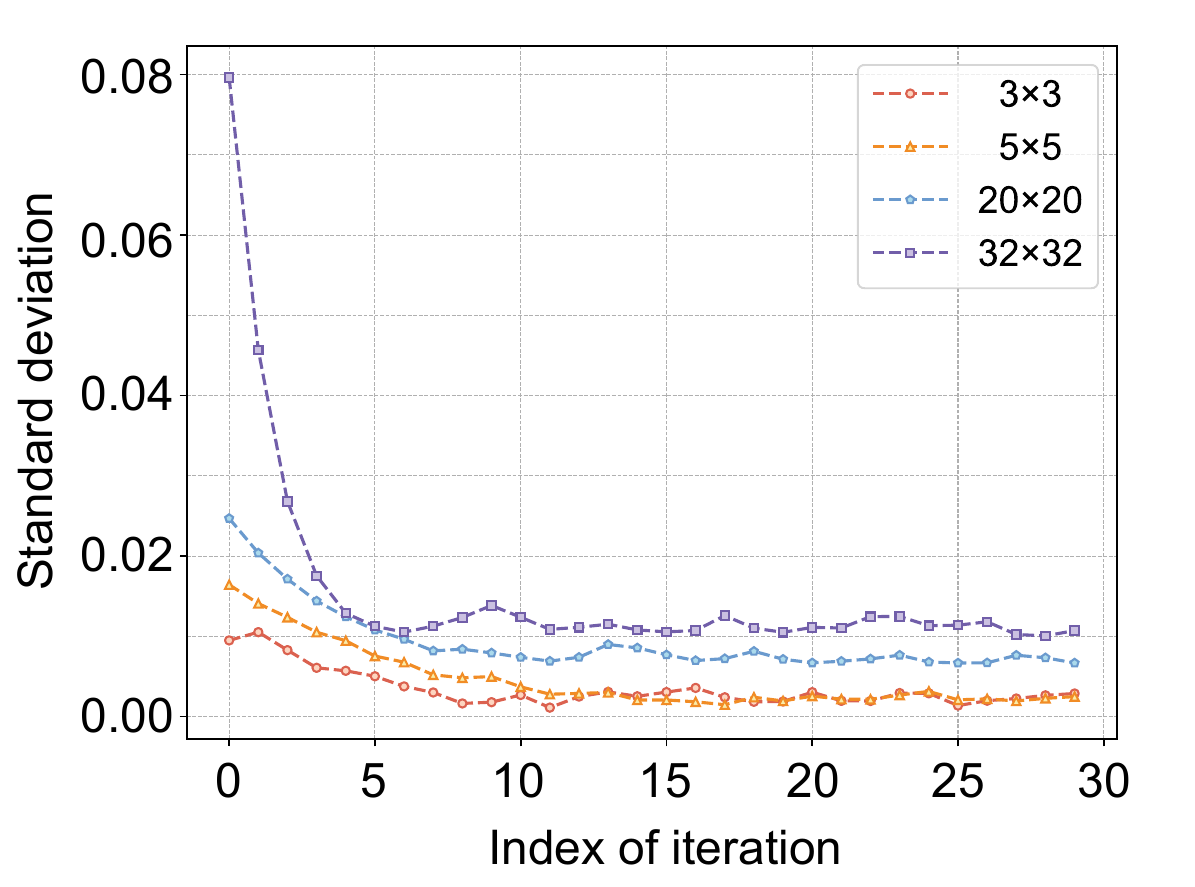}
\caption{Non-uniformity (standard deviation) with different array sizes. The data are the average results of 10 experiments. The error is one standard variance}\label{fig4}
\end{figure}

\begin{figure}
\centering\includegraphics[width=0.5\textwidth]{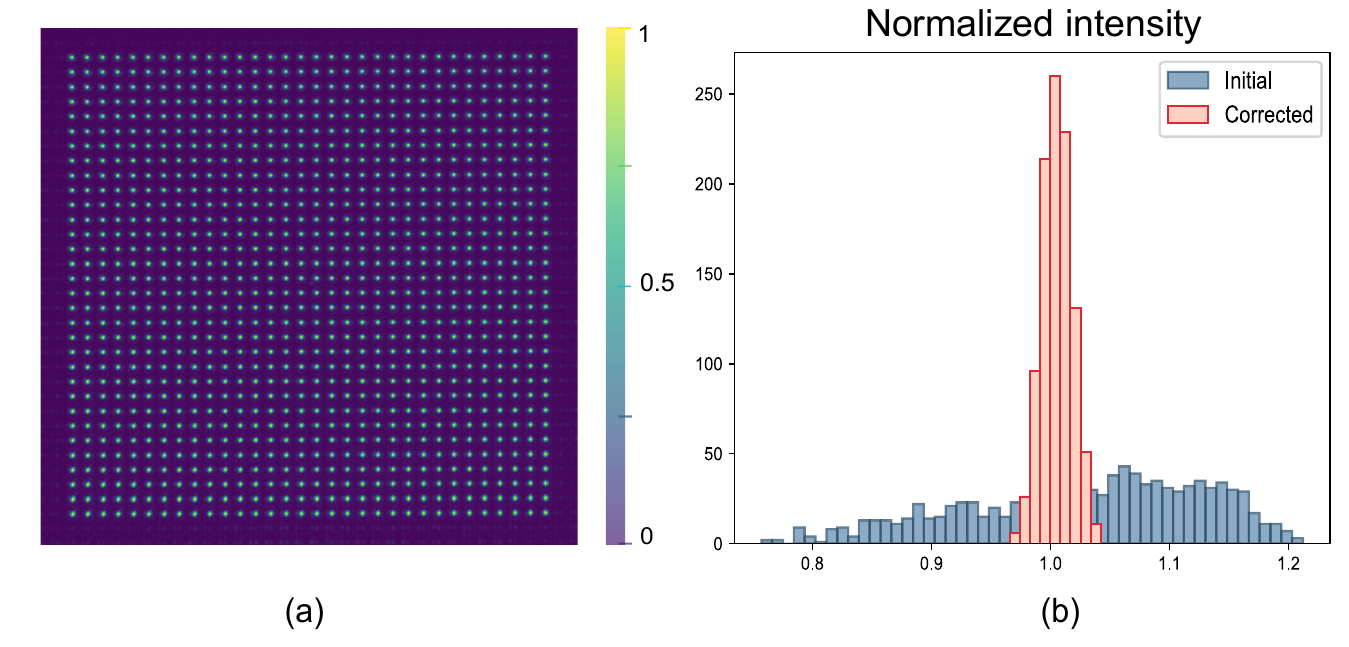}
\caption{Experimental results of 32$\times$32 tweezer array.(a) shows the experimental results of 32$\times$32 tweezers array after intensity equalization,the non-uniformity of this array is 1.006\%. (b) is a comparison of intensity distribution between the GSW algorithm and the feedback GSW algorithm. After feedback optimization, the strength equalization is significantly improved}\label{fig5}
\end{figure}

We evaluated the overall performance of the algorithm, and when the GSW algorithm was iterated 20 times, the algorithm results tended to be stable. As shown in Fig.\ref{fig4}, the initial standard deviation increases as the array size increases.And when the camera feedback loop was done 10 times, the actual uniformity results tended to be stable. 
From the experimental results, we can find that the test uniformity of the GSW algorithm without image feedback tends to deteriorate when the array size increases.To present the experimental results of intensity equalization, we analyzed the final array results of the 32×32 array in Fig.\ref{fig4}. Fig.\ref{fig5} (a) is a graph of normalized intensity data, which is the date acquired from Thorlabs' CMOS Camera Beam Profiler (BC106N - VIS/M).  Fig.\ref{fig5}(b) shows the statistical results of the tweezer array intensities. It can be seen from the results that the uniformity of the array has been significantly improved by the feedback intensity homogenization algorithm.

\maketitle
\section{Conclusion and outlook}
The neutral atomic array system is one of the most popular experimental systems in quantum computation and simulation. We built up this static tweezer system for trapping atomic arrays. To obtain an ideal single atomic array, the intensity equalization algorithm of the static tweezer array system is developed. We find that the fixed initial phase GSW algorithm can reduce the fluctuation of array uniformity and obtain better uniformity. Therefore, by several random phase GSW algorithms, the best initial phase pattern corresponding to the best homogenization result is selected. For the feedback GSW algorithm next, the selected initial phase pattern is used for better uniformity. The results show that this method can reduce the non-uniformity by about 1.1\%. On the other hand, the tweezers in the focal plane of the holographic tweezer array only occupy several pixels, so the intensity distribution obtained by the CCD camera must be a saw-tooth shape. These saw-tooth shapes will lead to inaccurate uniformity results. Therefore, we use two-dimensional Gaussian fit to obtain the intensity distribution. At the same time, two-dimensional Gaussian fit can also give out the information of spot size uniformity, and ellipticity of the tweezer trap that provides more judgment methods for further system optimization. The results show that the non-uniformity can below 1.1\% when the tweezer size is bigger than 1000. The realization of a high uniformity tweezer array is of great significance to the many-body physics based on the single atom array.

\maketitle
\section*{Acknowledgements}

This work was supported by National Key R\&D Program of China under Grant No. 2020YFA0309400, NNSFC under Grant No. 12222409, and the Key Research and Development Program of Shanxi Province (Grant No. 202101150101025).

\end{document}

\title{Supplementary Material for Coherent dressing of ground-state spin in a thermal atomic ensemble}

\author{Xin Jia}
\thanks{These authors contributed equally to this work.}
\affiliation{State Key Laboratory of Quantum Optics and Quantum Optics Devices, Institute of Opto-Electronics, Shanxi University, Taiyuan 030006, China}

\author{Weixin Liu}
\thanks{These authors contributed equally to this work.}
\affiliation{Department of Physics, Xinzhou Normal University, Xinzhou  034000, China}
\affiliation{Department of Physics, Chongqing University, Chongqing 401331, China}

\author{Shengxian Xiao}
\affiliation{Department of Physics, Chongqing University, Chongqing 401331, China}

\author{Zimo Zhang}
\affiliation{State Key Laboratory of Quantum Optics and Quantum Optics Devices, Institute of Opto-Electronics, Shanxi University, Taiyuan 030006, China}

\author{Bin Huang}
\affiliation{State Key Laboratory of Quantum Optics and Quantum Optics Devices, Institute of Opto-Electronics, Shanxi University, Taiyuan 030006, China}

\author{Tao Wang}
\email{tauwaang@cqu.edu.cn}
\affiliation{Department of Physics, Chongqing University, Chongqing 401331, China}

\author{Bing Chen}%
\email{bingchenphysics@hfut.edu.cn}
\affiliation{School of Physics, Hefei University of Technology, Hefei 230009, China}

\author{Heng Shen}%
\email{hengshen@sxu.edu.cn}
\affiliation{State Key Laboratory of Quantum Optics and Quantum Optics Devices, Institute of Opto-Electronics, Shanxi University, Taiyuan 030006, China}
\affiliation{Collaborative Innovation Center of Extreme Optics, Shanxi University, Taiyuan 030006, China}

\maketitle

\setcounter{equation}{0}
\setcounter{figure}{0}
\renewcommand{\theequation}{S\arabic{equation}}
\renewcommand{\thefigure}{S\arabic{figure}}
\renewcommand{\thetable}{S\Roman{table}}
\renewcommand{\bibnumfmt}[1]{[S#1]}
\renewcommand{\citenumfont}[1]{S#1}

\section{Theoretical Method}
In this section, we derive the probe transmission of coherent dressing of ground-states under the level diagram in Fig. 1(a), and then give a intuitive explanation to the EIT profiles. The relevant Hamiltonian in an optical rotating frame takes the form in the basis of bare states ($[|1\rangle, |2\rangle, |3\rangle, |e\rangle]^T$)

\begin{align}\label{1}
\hat{H}(t)=-\frac{\hbar}{2}\!
\begin{pmatrix}
0    &\Omega_0(1\!+\!e^{-2i\omega_0t})     &0    &\Omega_p \\
\Omega_0(1\!+\!e^{2i\omega_0t})  &2\Delta_0   &  \Omega_0(1\!+\!e^{-2i\omega_0t})    &0\\
0   &\Omega_0(1\!+\!e^{2i\omega_0t})  & 4\Delta_0  & \Omega_ce^{i\Delta t} \\
\Omega_p   &0     &  \Omega_ce^{-i\Delta t}   &2\Delta_p
\end{pmatrix} , \tag{S1}
\end{align}
where $\Delta=\Delta_c-\Delta_p+2\Delta_0$, and $\Omega_0$ is the Rabi frequency of RF magnetic field, $\Omega_p(c)$ is the Rabi frequencies of the probe (couple) laser; and $\Delta_0=\omega_0-\omega_L$ is the RF field detuning, $\Delta_{p(c)}=\omega_{p(c)}-\omega_{e1(e3)}$ is the detuning of probe (couple) laser.

The dynamic of the system is govern by the associated master equation 
\begin{equation}\label{2}
\dot{\rho}=-\frac{i}{\hbar}[\hat{H},\rho]+\mathcal{L}[\rho],  \tag{S2}
\end{equation}
where the Liouvillian term $\mathcal{L}[\rho]$ describes the relaxation process, and reads as
\begin{align}\label{3}
\mathcal{L}[\rho]=
\begin{pmatrix}
	\gamma_0(\rho_{22}\!-\!\rho_{11})\!+\!\frac{\Gamma}{3}\rho_{ee}   &-\!\frac{\gamma_0}{2}\rho_{12}
	&-\!\frac{\gamma_0}{2}\rho_{13}    &-\!\frac{\gamma_0+\Gamma}{2}\rho_{1e} \\
	-\frac{\gamma_0}{2}\rho_{21} &\gamma_0(\rho_{11}\!+\!\rho_{33}\!-\!2\rho_{22})\!+\frac{\Gamma}{3}\rho_{ee}  
	&-\!\frac{\gamma_0}{2}\rho_{23} 
	&-\!\frac{\gamma_0\!+\!\Gamma}{2}\rho_{2e}\\
	-\!\frac{\gamma_0}{2}\rho_{31} 
	&-\!\frac{\gamma_0}{2}\rho_{32}
	&	\gamma_0(\rho_{22}\!-\!\rho_{33})\!+\!\frac{\Gamma}{3}\rho_{ee}   &-\frac{\gamma_0\!+\!\Gamma}{2}\rho_{3e}\\
	-\!\frac{\gamma_0\!+\!\Gamma}{2}\rho_{e1}  &-\frac{\gamma_0\!+\!\Gamma}{2}\rho_{42} &-\frac{\gamma_0\!+\!\Gamma}{2}\rho_{e3} 
	&-\!\Gamma\rho_{ee}
\end{pmatrix}. \tag{S3}
\end{align}
Here, $\Gamma$ represent the spontaneous decay rate of excited state, while  $\gamma_0$ characterizes the spin depolarization of ground state, . 

The EIT profiles are obtained by detecting the probe laser transmission associated with the  imaginary component of optical coherence element, that is, $-$Im$\rho_{e1}$. For our model, the $\rho_{e1}$ here is the average steady-state density matrix element, corresponding to the transition $|1\rangle\leftrightarrow|e\rangle$. Building upon this model, we exemplify a numerical result compared with experimental data in Fig.\ref{figs1}. We can see that the theoretical results are in a good accordance with experimental data, exhibiting a Mollow-triplet like profile with five equally distributed resonance peaks. 

\begin{figure}[htp]
	\centering
	\includegraphics[width=0.7\linewidth]{FigSS1.pdf}
	\caption{Numerical (a) and experimental (b) results of the EIT spectra with $\omega_0/2\pi=1.955$ kHz and $\Omega_0/2\pi=0.9$ kHz.}\label{figs1}%
\end{figure}

\begin{figure}[htp]
  \centering
   \includegraphics[width=0.7\linewidth]{FigS1.pdf}
   \caption{The energy level diagram of the four-level system in the bare picture and the ground-state dressed picture.}\label{figs2}%
\end{figure}

\section{Dressed state picture}
Here we give an intuitive understanding of the EIT spectrum of our system. At first, the counter-rotation terms that proportion to  $e^{\pm 2i\omega_0t}$ remain to be kept in the Hamiltonian Eq.(\ref{1}) given $\Omega_0$ is comparable to $\omega_0$. This results in the anomalous EIT peaks shown in Fig.2b in main text, which gradually obstruct the other peaks as increasing $\Omega_0$. However, for small $\Omega_0$ we consider in this section, we can neglect the counter-rotation terms safely.  Thus those EIT peaks can be easily understood in the ground-state dressed picture. By diagonalizing partly Hamiltonian, i.e., the first 3$\times$3 matrix, we get the RF-dressed eigenstates of the system which consists of the three bare ground states interacting with the RF magnetic field, and the corresponding dressed eigenenergies can be expressed by 
\begin{align}\label{4}
&E_0/\hbar=-\Delta_0,\tag{S4a}\\
&E_{\pm}/\hbar=-\Delta_0\pm\sqrt{\Delta_0^2+\Omega_0^2/2}.\tag{S4b}
\end{align}
In such a partly dressed picture, the bare picturre Hamiltonian Eq.(\ref{1}) changed into
\begin{align}\label{5}
	\hat{H}_{d}(t)=-\frac{\hbar}{2}
	\begin{pmatrix}
		 2\Delta_0-2\sqrt{\Delta^2_0+\Omega^2_0/2}& 0 & 0  &\Omega^{'}_p+\Omega^{'}_c e^{i\Delta t}\\
		0& 2\Delta_0  & 0&  \Omega^{''}_p+\Omega^{''}_c e^{i\Delta t} \\
		0&0   &  2\Delta_0+2\sqrt{\Delta^2_0+\Omega^2_0/2}  & \Omega^{'''}_p+\Omega^{'''}_ce^{i\Delta t} \\
		\Omega^{'}_p+\Omega^{'}_c e^{-i\Delta t}&\Omega^{''}_p+\Omega^{''}_c e^{-i\Delta t}     &\Omega^{'''}_p+\Omega^{'''}_c e^{-i\Delta t}   &2\Delta_p
	\end{pmatrix} , \tag{S5}
\end{align}
where $\Omega^{'}_p=\frac{\Omega^2_0\Omega_p}{2(2\Delta^2_0+\Omega^2_0+\sqrt{2}\Delta_0\sqrt{2\Delta^2_0+\Omega^2_0})}$, $\Omega^{'}_c=\frac{\Omega^2_0\Omega_c}{2(2\Delta^2_0+\Omega^2_0-\sqrt{2}\Delta_0\sqrt{2\Delta^2_0+\Omega^2_0})}$, $\Omega^{''}_p=-\frac{\Omega_0\Omega_p}{\sqrt{2}\sqrt{2\Delta^2_0+\Omega^2_0}}$, $\Omega^{''}_c=\frac{\Omega_0\Omega_c}{\sqrt{2}\sqrt{2\Delta^2_0+\Omega^2_0}}$, $\Omega^{'''}_p=\frac{\Omega^2_0\Omega_p}{2(2\Delta^2_0+\Omega^2_0-\sqrt{2}\Delta_0\sqrt{2\Delta^2_0+\Omega^2_0})}$, $\Omega^{'''}_c=\frac{\Omega^2_0\Omega_c}{2(2\Delta^2_0+\Omega^2_0+\sqrt{2}\Delta_0\sqrt{2\Delta^2_0+\Omega^2_0})}$.

In the Hamiltonian Eq.(\ref{5}) we can see that three RF-dressed states labeled $|0\rangle$, $|\pm\rangle$ and the bare excited state constitute a new four-level system, and each dressed state coupled the excited state with both the probe and coupling field, corresponding to the modified probe and couple Rabi frequency, respectively. As a consequence, this new four-level system includes five different $\Lambda$-type three-level subsystems as shown in Fig.\ref{figs2}. For example, one of the five subsystems is $|-\rangle-|e\rangle-|0\rangle$, and the corresponding Hamiltonian of which is 
\begin{align}\label{6}
	\hat{H}^{'}_d(t)=-\frac{\hbar}{2}
	\begin{pmatrix}
		2\Delta_0-2\sqrt{\Delta^2_0+\Omega^2_0/2}& 0  &\Omega^{'}_p+\Omega^{'}_c e^{i\Delta t}\\
		0& 2\Delta_0 &\Omega^{''}_p+\Omega^{''}_c e^{i\Delta t} \\
		\Omega^{'}_p+\Omega^{'}_c e^{-i\Delta t}&\Omega^{''}_p+\Omega^{''}_c e^{-i\Delta t}    &2\Delta_p
	\end{pmatrix}.\tag{S6}
\end{align}
We can see from the Hamiltonian Eq.(\ref{6}) that the three-level subsystem $|-\rangle-|e\rangle-|0\rangle$ implies two physical processes that can produce EIT, i.e.,
\begin{align}\label{7}
	\hat{H}^{'}_{1d}(t)=-\frac{\hbar}{2}
\begin{pmatrix}
		2\Delta_0-2\sqrt{\Delta^2_0+\Omega^2_0/2}& 0  &\Omega^{'}_p\\
		0& 2\Delta_0 &\Omega^{''}_c e^{i\Delta t} \\
		\Omega^{'}_p&\Omega^{''}_c e^{-i\Delta t}   &2\Delta_p
\end{pmatrix},
\;
\hat{H}^{'}_{2d}(t)=-\frac{\hbar}{2}
\begin{pmatrix}
		2\Delta_0-2\sqrt{\Delta^2_0+\Omega^2_0/2}& 0  &\Omega^{'}_c e^{i\Delta t}\\
		0& 2\Delta_0 &\Omega^{''}_p \\
		\Omega^{'}_c e^{-i\Delta t}&\Omega^{''}_p    &2\Delta_p
\end{pmatrix}. \tag{S7}
\end{align}
Now we first focus on the process $\hat{H}^{'}_{1d}$. Using the unitary operator \textcolor{red}{$\hat{U}=|-\rangle\langle-|+e^{i\Delta t}|0\rangle\langle0|+|+\rangle\langle+|$} to transform $\hat{H}^{'}_{1d}$ into a time-independent form
\begin{align}\label{8}
	\hat{H}^{'}_{1^{'}d}(t)=-\frac{\hbar}{2}
	\begin{pmatrix}
		2\Delta_0-2\sqrt{\Delta^2_0+\Omega^2_0/2}& 0  &\Omega^{'}_p\\
		0& 2\Delta_p- 2\Delta_c-2\Delta_0&\Omega^{''}_c \\
		\Omega^{'}_p&\Omega^{''}_c    &2\Delta_p
	\end{pmatrix}. \tag{S8}
\end{align}
Thus according to the "two-photon resonance", we can get the condition of EIT from Eq.(\ref{8}), that is, $\Delta_p=\Delta_c+2\Delta_0-\sqrt{\Delta^2_0+\Omega^2_0/2}$. Similarly, from the Hamiltonian $\hat{H}^{'}_{2d}$ we can get another resonance condition $\Delta_p=\Delta_c+2\Delta_0+\sqrt{\Delta^2_0+\Omega^2_0/2}$. 

\textcolor{red}{In the same manner, we now list all the subsystems which corresponding respectively to $|-\rangle-|e\rangle-|-\rangle$ (which be equivalent to $|0\rangle-|e\rangle-|0\rangle$, $|+\rangle-|e\rangle-|+\rangle$ because they just differ in the probe detuning and coupling detuning both shifting a same value), $|-\rangle-|e\rangle-|0\rangle$ (which be equivalent to $|0\rangle-|e\rangle-|+\rangle$),  $|-\rangle-|e\rangle-|+\rangle$, $|+\rangle-|e\rangle-|0\rangle$ (which be equivalent to $|0\rangle-|e\rangle-|-\rangle$), $|+\rangle-|e\rangle-|-\rangle$, then we can get the resonance conditions of all the different five EIT spectra, that are}
\begin{align}\label{9}
&\Delta_p=\Delta_c+2\Delta_0,\tag{S9a}\\
&\Delta_p=\Delta_c+2\Delta_0-\sqrt{\Delta_0^2+\Omega_0^2/2},\tag{S9b}\\
&\Delta_p=\Delta_c+2\Delta_0+\sqrt{\Delta_0^2+\Omega_0^2/2},\tag{S9c}\\
&\Delta_p=\Delta_c+2\Delta_0-2\sqrt{\Delta_0^2+\Omega_0^2/2},\tag{S9d}\\
&\Delta_p=\Delta_c+2\Delta_0+2\sqrt{\Delta_0^2+\Omega_0^2/2},\tag{S9e}
\end{align}
For $\Delta_c=0$ and $\Delta_0=0$, these conditions imply that the positions of EIT peaks increasing linearly with $\Omega_0$, as shown in the Fig.2b in the main text.  For $\Delta_0\neq0$, we compare the EIT conditions Eq.(S9) with the numerical results in the Fig.\ref{figs3}, and we can see that the five EIT conditions given by Eq.(S9) agrees well with the numerical simulation.

\begin{figure}
  \centering
  \includegraphics[width=0.7\linewidth]{Figs2.pdf}
  \caption{The numerical (a) and experimental (b) EIT spectra varied with $\Delta_p$ and $\Delta_0$ under $\Omega_0/2\pi=$1kHz. The dashed red lines are plotted according to Eq. (S9).}\label{figs3}%
\end{figure}

\begin{figure}
\centering
\includegraphics[width=0.5\textwidth]{FigS3.pdf}
\caption{\label{numerical} The theoretical upper boundary of the frequency difference $\Omega_0/2\pi$ of two split lines $\delta_{0,-1}$ as a function of the Zeeman level splitting $\omega_0/2\pi$. The blue dot is the result of theoretical calculation. The dashed red line is the fitting curve.}
\end{figure}

In addition, we theoretically calculate the upper boundary of RF signal intensity $\Omega_0/2\pi$ to which the atomic system can response linearly (Fig. \ref{numerical}). The theoretical upper boundary $\Omega_0/2\pi$ is linear with Zeeman level splitting $\omega_0/2\pi$. Comparing with Fig. 3 (b) in the main text, there is difference between the experimental and the theoretical results as the Zeeman level splitting is small. For the Zeeman splitting $\omega_0/2\pi < 1.5$ kHz, the experimental results deviate from the fitting line. Under small magnetic fields, the RF magnetic field not only facilitates atomic population evolution but also influences the Zeeman level splitting. In contrast, the role of RF magnetic field remains singular in the theoretical calculation.

\section{Calibration of the RF coil}

\begin{figure}
\centering
\includegraphics[width=0.5\textwidth]{FigS4.pdf}
\caption{\label{Exp} The measured RF magnetic field amplitude $B^{V_{pp}}_{\text{RF}}$ for different applied voltage $V_{pp}$ at 1.955 kHz. "pp" means peak-to-peak value.}
\end{figure}

In the experiment, a pair of square Helmholtz coils aligned parallel to each other and perpendicular to the DC bias magnetic field is used to generate the RF magnetic field $B_{\text{RF}}^{\text{Vpp}} \sin(\omega t)$ (Stanford Research Systems SRS DS345). To calibrate the RF magnetic field amplitude generated by different RF set voltages from the signal generator, a pick-up coil with $N_{\omega} = 30$ turns and a diameter of 8.73 mm is placed at the center of the RF magnetic field, which is the location of the atomic vapor cell. Due to the oscillating RF magnetic field, a magnetic flux is induced in the pick-up coil, generating an electromotive force (EMF) $\epsilon_{\omega}$. The pick-up coil is connected to a lock-in amplifier (Zurich Instruments HF2LI), which demodulates the EMF to obtain the potential $U_m = \frac{U_{\omega}}{\sqrt{2}}$, where the amplitude ${U_{\omega}}$ is defined from the time dependent voltage $U(t)={U_{\omega}e^{i{\omega}t}}$ and $\sqrt{2}$ comes from the fact that the demodulated signal from the lock-in amplifier is the root mean square (RMS) value. The RF magnetic field strength $B_{\text{RF}}^{\text{Vpp}}$ is given by \cite{HS thesis,Bao}:
\begin{equation}
B_{\text{RF}}^{\text{Vpp}} = \frac{(1 + \frac{Z_{\text{coil}}}{R_m}) U_{\omega}}{N_{\omega} A_{\text{coil}}\omega}
\end{equation}
where $(1+\frac{Z_{\text{coil}}}{R_m}) U_m$ is the EMF $\epsilon_{\omega}$, $R_m = 50 \, \Omega$ is the measurement impedance of the lock-in amplifier, and $Z_{\text{coil}} = \frac{(R + i \omega L)}{i \omega C} / {(R + i \omega L + \frac{1}{i\omega C})}$ is the impedance of the pick-up coil, with $R$, $L$, and $C$ being the resistance, inductance, and capacitance of the pick-up coil, respectively. $A_{\text{coil}}$ is the cross-sectional area of the pick-up coil. Fig.\ref{Exp} shows that at a frequency of 1.955 kHz, the RF magnetic field amplitude $B_{\text{RF}}^{\text{Vpp}}$ exhibits a linear response to the applied voltage.

\section{Additional Data}
In Fig.\ref{1M} we plot the EIT spectra with $\omega_0/2\pi=1.003241$ MHz with $F=1\to F'=1$ transition. The inhomogeneity of the magnetic field causes the broaden resonance peaks.
\begin{figure}
\centering
\includegraphics[width=0.5\textwidth]{FigS6.pdf}
\caption{\label{1M} EIT spectra with $\omega_0/2\pi=1.003241$ MHz with $\Delta_0=0$.}
\end{figure}